\documentclass[traditabstract,letter]{aa}
\usepackage{amssymb,amsmath,amsfonts}
\usepackage{ulem}             
\usepackage{pgf,tikz}
\usepackage{natbib}
\usepackage{hyperref}
\hypersetup{colorlinks=true,
            citecolor=blue,
            linkcolor=blue}
 \usepackage{lipsum}

\def\apjl{Astrophys.\ J. }
\def\apjs{Astrophys.\ J.\ (Supp.) }
\def\aap{Astron.\ Astrophys. }

\newcommand{\be}{\begin{equation}}
\newcommand{\ee}{\end{equation}}

\DeclareMathOperator{\atan}{atan}

 {\everymath{\displaystyle\everymath{}}\array}%
 {\endarray}

\newcommand\Ha{{\alpha}}
\newcommand\gO{{\cal O} }
\newcommand\eps{{\varepsilon} }

\begin{document}

\title{Detection of   co-orbital planets by combining \\ transit and radial-velocity measurements}

\titlerunning{Detection of co-orbital planets combining transits and radial-velocities}

\author{
A. Leleu\inst{1,2}
\and P. Robutel\inst{2}
\and A. C. M Correia\inst{2,3}
\and J. Lillo-Box\inst{4}
}

\institute{
CHEOPS fellow, Physikalisches Institut, Universitaet Bern, CH-3012 Bern
  \and 
IMCCE, Observatoire de Paris - PSL Research University, UPMC Univ. Paris 06, Univ. Lille 1, CNRS,
77 Avenue Denfert-Rochereau, 75014 Paris, France
\and
CIDMA, Departamento de F\'isica, Universidade de Aveiro, Campus de
Santiago, 3810-193 Aveiro, Portugal
  \and 
European Southern Observatory, Alonso de Cordova 3107, Vitacura Casilla 19001, Santiago 19, Chile 
}

\date{\today}

\abstract{
Co-orbital planets have not yet been discovered, although they constitute a frequent by-product of planetary formation and evolution models.
This lack  may be due to observational biases, since the main detection methods are unable to spot co-orbital companions when they are small or near the Lagrangian equilibrium points.
However, for a system with one known transiting planet (with mass $m_1$), we can detect a co-orbital companion (with mass $m_2$) by combining the time of mid-transit with the radial-velocity data of the star.
Here, we propose a simple method that allows the detection of co-orbital companions, valid for eccentric orbits, that relies on a single parameter $\Ha$, which is proportional to the mass ratio $m_2/m_1$.
Therefore, when $\Ha$ is statistically different from zero, we have a strong candidate to harbour a co-orbital companion.
We also discuss the relevance of false positives generated by different planetary configurations.
}


 \maketitle

%

\section{Introduction}

Co-orbital planets consist of two planets with masses $m_1$ and $m_2$ orbiting with the same mean motion a central star with mass $m_0$. 
In the quasi-circular case, as long as the mutual inclination remains smaller than a few tens of degrees, the only stable configurations are the Trojan (like Jupiter's trojans) and the Horseshoe (like Saturn's satellites Janus and Epimetheus). 
Stable Trojan configurations arise for $(m_1+m_2)/m_0 \lesssim 4\times 10^{-2} $ \citep{Ga1843}, and  Horseshoe configurations for $(m_1+m_2)/m_0 \lesssim 2 \times 10^{-4}$ \citep{LauCha2002}. 
We note that, at least when no dissipation is involved, the stability of a given configuration does not depend much on the mass distribution between $m_1$ and $m_2$. 

Co-orbital bodies are common in the solar system and are also a natural output of planetary formation models \citep{CreNe2008,CreNe2009}.
However, so far none have been found in exoplanetary systems,
likely owing to the difficulty  in detecting them.
For small eccentricities, there is a degeneracy between the signal induced by two co-orbital planets and a single planet in an eccentric orbit or two planets in a 2:1 mean motion resonance \citep[e.g.][]{GiuBe2012}.
In favourable conditions, both co-orbital planets can eventually be observed transiting in front of the star, but this requires two large radii and small mutual inclination. 
A search for co-orbital planets was made using the {\it Kepler Spacecraft}\footnote{http://kepler.nasa.gov/} data, but none were found \citep{Ja2013,Fa2014}. 
We hence conclude that co-orbitals are rare in packed multi-planetary systems (like those  discovered by {\it Kepler}), that they are not coplanar, or that one co-orbital is much smaller than the other.
For larger semi-major axes, we expect that at least one of the co-orbitals cannot be observed transiting.
When the libration amplitude of the resonant angle is detectable (either by transit-time variations or with radial-velocity modulations), we can still infer the presence of both planets \citep{LauCha2002,FoHo2007}.
These effects have not been detected so far, at least not with  sufficient precision to rule out other scenarios. However,
we cannot conclude that no co-orbitals are present in the observed systems: transit timing variation (TTV) and radial-velocity methods will both miss a co-orbital companion if the amplitude of libration is not large enough or if its period is too long.

\citet{FoGa2006} noticed that for a single planet in a circular orbit, the time of mid-transit coincides with the instant where the radial-velocity reaches its mean value.
However, if the planet that is transiting has a co-orbital companion located at one of its Lagrangian points, there is a time shift $\Delta T$ between the mid-transit and the mean radial-velocity, that depends on the properties of the co-orbital companion.
Therefore, when we combine transit and radial-velocity measurements, it is possible to infer the presence of a co-orbital companion.
This method was developed for circular orbits and for a companion at the exact Lagrangian point (without libration). Although it remains valid for small libration amplitudes (which would just slightly modify the determined mass), co-orbital exoplanets can be stable for any amplitude of libration.
Moreover, for a single transiting planet in a slightly eccentric orbit, we can also observe the same time shift $\Delta T$, without requiring the presence of a co-orbital companion.

In this Letter, we generalise the work by \cite{FoGa2006} to eccentric planets in any Trojan or Horseshoe configuration (any libration amplitude).
When a planet is simultaneously observed through the transit and radial-velocity techniques,
we propose a simple method for  detecting the presence of a co-orbital companion that relies on a single dimensionless parameter $\Ha \propto m_2/m_1$.
Therefore, when $\Ha$ is statistically different from zero, we have a strong candidate to harbour a co-orbital companion and we get an estimation of its mass.
Moreover, if the secondary eclipse of the transiting planet is also observed, our method further constrains the uncertainty in $\Ha$. 
We also discuss the possibility of false positive detections due to other effects. 

\section{Radial-velocity}
\label{sec:RV}
 
 \begin{figure}
\begin{center}
\includegraphics[width=0.78\linewidth]{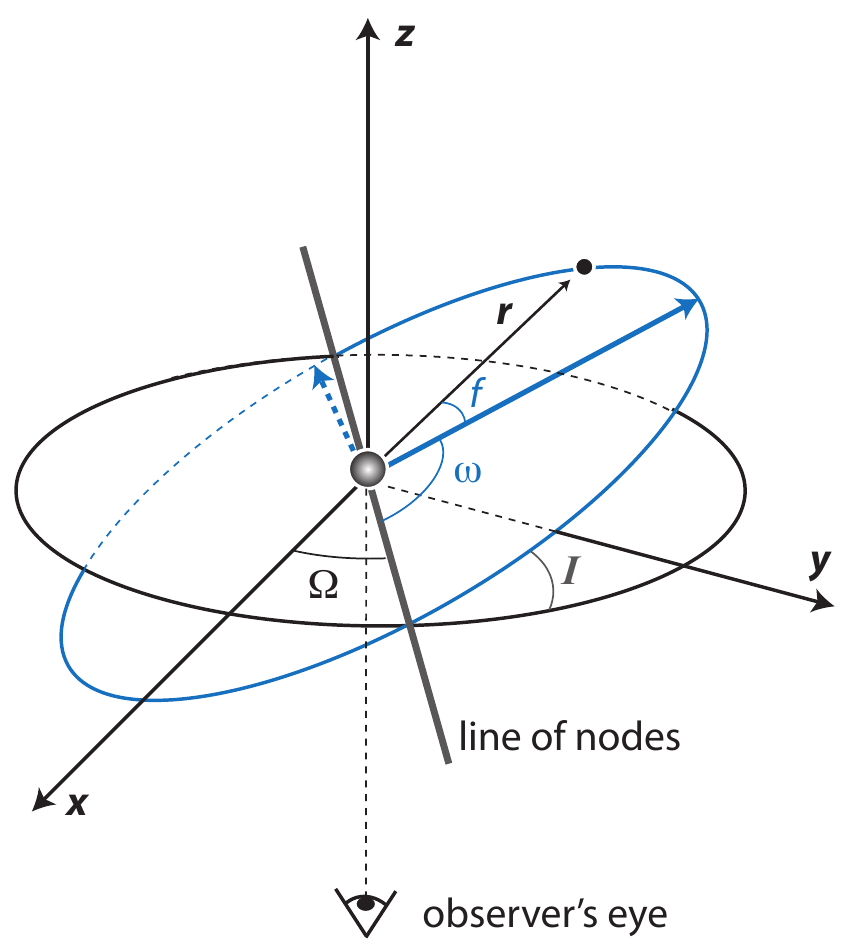}
\caption{\label{fig:stabzr} Reference angles for the orbit of a given planet with respect to an arbitrary frame Oxyz, where O is the centre of the star and z the line of sight.}
\end{center}
\end{figure}

In a reference frame where the $z$-axis  coincides with the observer's line of sight (Fig.\,\ref{fig:stabzr}), the radial-velocity of the star 
induced by the planet $k$ with mass $m_k$ is given by \citep{MuCo2010}
%
  \begin{equation}
v_k = 
- \frac{K_k}{\sqrt{1-e_k^2}} (\cos \ell_k + e_k \cos \omega_k ) \ ,
   \label{eq:RVj}
   \end{equation}
with 
  \begin{equation}
K_k = \frac{m_k}{M} n_k a_k \sin I_k \ ,
\quad \mathrm{and} \quad
\ell_k = \omega_k + f_k \ , 
   \label{eq:RVj2}
   \end{equation}
where $M = m_0 + \sum_k m_k$, $a$ is the semi-major axis, $n$ is the mean motion, $e$ is the eccentricity, $I$ is the inclination angle between the plane of the sky and the orbital plane, $\omega$ is the argument of the pericentre, and $f$ is the true anomaly. 
%
   %
%

For small eccentricities, we can simplify $v_k$ by expanding $\cos  \ell_k$ in powers of $e_k$ \citep{1999ssd}
  \begin{equation}
\cos \ell_k = \cos \lambda_k + e_k \cos (2 \lambda_k - \omega_k) - e_k \cos \omega_k + \gO(e_k^2) \ ,
   \label{eq:murray}
   \end{equation}
where $\lambda_k = n_k t + \varphi_k$, and $\varphi_k$ is a phase angle.
At first order in eccentricity, the radial-velocity induced by a single planet on a Keplerian orbit is thus of the form
   \begin{equation}
   \begin{aligned}
v_{k} =  & A_{k} \cos n_k t+ B_{k} \sin n_k t\\
& + C_{k} \cos 2 n_k t + D_{k} \sin 2 n_k t \ ,
\end{aligned}  
   \label{eq:RVeq}
   \end{equation}
with               
   \begin{equation}
      \begin{aligned}
         A_{k} &=- K_{k} \cos\,\varphi_{k} \, ,  &C_{k} &=- e_{k}K_k \cos\, (2\varphi_{k}-\omega_{k}) \, ,\\
        B_{k}&= K_{k} \sin\,\varphi_{k}\, ,   &D_{k}&= e_{k}K_{k} \sin\, (2\varphi_{k}-\omega_{k}) \,  . \\
\end{aligned}
   \label{eq:RV1val}
   \end{equation}   
If we sum the contribution of two planets on Keplerian orbits, the total radial-velocity of the star becomes
     \begin{equation}
      \begin{aligned}
  v =  \gamma + v_1 + v_2 \, ,
  \end{aligned}
    \label{eq:RVsum}
   \end{equation} 
   where $\gamma$ is the velocity of the system's barycentre.
In the co-orbital quasi-circular case, the semi-major axes of the planets librate around their mean value $\bar a$ with a frequency $\nu \propto \sqrt{\mu} n$, where $\mu = (m_1+m_2)/M$ and $n$ is the mean-motion associated with $\bar a$.
The amplitude of the libration goes from $0$ at the Lagrangian equilibrium up to $\gO(\sqrt{\mu})$ in the tadpole domain, and to $\gO(\mu^{1/3})$ in the horseshoe domain \citep{1977erdi,RoPo2013}. We note that horseshoe co-orbitals are stable only for $\mu$ lower than $\approx 2\times 10^{-4}$. For a pair of co-orbital planets we hence have $n_1-n_2=\gO(\mu^{\beta},  e_k^2),$ where $ \beta \geq 1/2$ for tadpole co-orbitals and $1/2 \geq \beta \geq 1/3$ for the horseshoe configuration.

There are two possible scenarios 
for which we can consider that $n_1= n_2 = n$: \\
1) when the time span is short with respect to the libration frequency $\nu$ and we do not have the frequency resolution to distinguish $n_1$ from $n_2$; \\
2) when the time span is longer than $2\pi/\nu$, and the harmonics of the radial-velocity signal are located at $p n + q \nu$ with $(p,q)\in \mathbb{Z}^2$. The harmonics for $q \neq 0$ have larger amplitudes if the co-orbitals librate with a large amplitude and if their masses are similar. 
If we can distinguish the effect of the libration in the radial-velocity signal, we can identify co-orbitals from radial-velocity alone \citep[see][]{LeRoCo2015}. 
If not, the assumption $n_1=n_2=n$ holds, and 
the mean longitudes 
simply read
\begin{equation}
\lambda_k = n t + \varphi_k + \gO(\mu^{\beta},  e_k^2) \, . \label{eq:longitude}
\end{equation}

For the radial-velocity induced by two co-orbitals, we hence sum cosines that have the same frequency. At order one in the eccentricities, we obtain an expression which is equivalent to (\ref{eq:RVeq}),
   \begin{equation}
   \begin{aligned}
v  =   \gamma & + A \cos n t+ B \sin n t\\
& + C \cos 2 n t + D \sin 2 n t \ ,
\end{aligned}  
   \label{eq:RVeqex}
   \end{equation}
with $A =A_1+A_2$, and similar expressions for $B$, $C$, and $D$. 
The radial-velocity induced by two co-orbitals is thus equivalent to the radial-velocity of a single planet on a Keplerian orbit with mean motion $n$, and orbital parameters given by
   \begin{equation}
      \begin{aligned}
        K &= \sqrt{A^2+B^2}\, ,  & e & =\frac{\sqrt{C^2+D^2}}{\sqrt{A^2+B^2}} \, ,\\
        \varphi &= -\atan \frac{B}{A}\, ,  & \omega & =  -2\atan\frac{B}{A} +\atan\frac{D}{C}  \, .\\
\end{aligned}
   \label{eq:RVeqval}
   \end{equation}  
These expressions are similar to those obtained by \citet{GiuBe2012}.
We note that this equivalence is broken at order $2$ in eccentricity: the next term in the expansion (\ref{eq:RVeqex}) is $E \cos 3 n t+F\sin 3nt$.
In the single planet case, we have 
\begin{equation}
\sqrt{E^2+F^2}=\frac{9}{8}\frac{C^2+D^2}{\sqrt{A^2+B^2}}+ \gO(e^4) \, ,
   \label{eq:A3}
\end{equation}
which is only also satisfied for eccentric co-orbitals for very specific values of the orbital parameters ($\lambda_1-\lambda_2=\omega_1-\omega_2$ and $e_1=e_2$). 
Therefore, in most cases, if we can determine $\sqrt{E^2+F^2}$, we can solve the degeneracy between a single planet and two co-orbitals.

\section{Time of mid-transit}

We now assume that the planet with mass $m_1$ is also observed transiting in front of the star. 
We consider that the planet transits when its centre of mass passes through the cone of light (we do not consider grazing eclipses because of the difficulty in estimating the time of mid-transit).
%
For simplicity, we set the origin of the time $t=0$ as the time of mid-transit.
%
  %
The true longitude $\ell_1$ of mid-transit is \citep{Wi2011}
\begin{equation}
\cos \ell_1 = e_1 \cos \omega_1 \cot^2 I_1 \, .
 \label{eq:trans4}
\end{equation}
The inclination $I_1$ has to be close to $\pi/2$ because the planet is transiting.
Denoting $I'_1=\pi/2-I_1$, we have that $e_1 \cos \omega_1 \cot^2 I_1 = \gO(e_1 I_1'^2)$, which is a negligible quantity. 
We thus conclude that for $t=0$,
\begin{equation}
\ell_1 = - \frac{\pi}{2} + \gO(e_1 I_1'^2) \, . \label{time0}
\end{equation}
%
%
We can now express the phase angles $\varphi_k$, involved in expressions (\ref{eq:RV1val}) and (\ref{eq:longitude}), in terms of $e_1$ and $\omega_1$. 
Since
   \begin{equation}
\lambda_1 = n t + \varphi_1 = \ell_1 - 2 e_1 \sin (\ell_1-\omega_1) + \gO(e_1^2)\, ,
   \label{eq:trans7}
   \end{equation}
it turns out that (using $t=0$)
   \begin{equation}
\varphi_1= - \frac{\pi}{2} + 2 e_1 \cos \omega_1 + \gO(e^2,e_1 I_1'^2)\, . 
   \label{eq:trans8}
   \end{equation}
For moderate mutual inclination and at order one in eccentricity we additionally have \citep{LeRoCo2015}
    \begin{equation}
    \begin{aligned}
\varphi_2 & = \varphi_1 + \zeta+ \gO (\mu,e^2,e\sqrt{\mu}) \ ,
        \end{aligned} 
   \label{eq:coorb}
   \end{equation}
where $\zeta = \lambda_2 - \lambda_1$ is the resonant angle. 
If we cannot see the impact of the evolution of $\zeta$ in the observational data, either because its amplitude of libration is negligible or because the libration is slow with respect to the time span of the measurements, we can consider $\zeta$ to be constant. 

\section{Radial-velocity and transit}
\label{sec:RVT}

In section~\ref{sec:RV}, we saw that, at first order in $e_k$, the radial-velocity induced by a pair of co-orbital planets is equivalent to that of a single planet. 
However, the phase angle $\varphi_1$ of the observed planet can be constrained by the transit event (Eq.\,(\ref{eq:trans8})).
Thus, assuming that we are able to measure the instant of mid-transit for the planet with mass $m_1$, we can replace the phase angles (\ref{eq:trans8}) and (\ref{eq:coorb}) in the expression of the radial-velocity (\ref{eq:RVeqex}) to obtain
\begin{equation}
\begin{aligned}
A &= - 2 K_1 k_1 - K_2 \left( \sin\,\zeta + 2 k_1 \cos\,\zeta \right) \  , \\
B &=  - K_1 - K_2 \left( \cos\,\zeta - 2 k_1 \sin\,\zeta \right) \  ,\\
C &= K_1 k_1+ K_2\left(  k_2 \cos\,2\zeta + h_2 \sin\,2\zeta \right)\ ,\\
D &= K_1 h_1+ K_2 \left( h_2 \cos\,2 \zeta - k_2 \sin\,2\zeta \right)\ ,
\end{aligned}  
 \label{eq:coef_vr1}
   \end{equation}
where $k_k=e_k \cos \omega_k$ and $h_k=e_k \sin \omega_k$. 

A striking result is that  the quantity
\begin{equation}
\begin{aligned}
A+2C = - K_2 \big( & \sin\,\zeta + 2 k_1 \cos\,\zeta  \\
      & - 2k_2 \cos\,2\zeta - 2h_2 \sin\,2\zeta \big) \, 
\end{aligned}  
 \label{eq:A2C}
   \end{equation}
is different from $0$ only if $K_2 \neq 0$, that is only  if the transiting planet $m_1$ has a co-orbital companion of mass $m_2$.
%
Therefore, the estimation of this quantity provides us invaluable information on the presence of a co-orbital companion to the transiting planet.

\section{Detection methods}
\label{sec:DM}

We assume that we are observing a star with a transiting planet, and that we are able to determine  the orbital period ($2\pi/n$) and the instant of mid-transit with a very high level of  precision.
We assume that radial-velocity data are also available for this star, and are consistent with the signal induced by a single planet on a slightly eccentric Keplerian orbit  (Eq.\,(\ref{eq:RVeqex})).  

Setting $t=0$ at the time of mid-transit, we propose a  fit to the radial-velocity data with the following function:
\begin{equation}
\begin{aligned}
v (t) =  \gamma + K \big[&(\Ha-2c) \cos n t  - \sin n t \\
       &+ c \cos 2 n t + d \sin 2 n t \big] \, .
\end{aligned}  
   \label{eq:RVf}
   \end{equation}
The parameters to fit correspond to $\gamma$, $K=-B$, $c = C/K$, $d = D/K$, and $\Ha=(A+2C)/K$. 
We fix $n$ because it is usually obtained from the transit measurements with better precision.
The dimensionless parameter $\Ha$ is proportional to the mass ratio $m_2/m_1$ (Eq.\,(\ref{eq:A2C})).
Whenever $\Ha$ is statistically different from zero, the system is thus a strong candidate to host a co-orbital companion. 


In general\footnote{Except when $\sin \zeta$ tends to $0$. However, this cannot happen when the sum of the mass of the co-orbital is higher than $10^{-3}$ the mass of the star, for stability reasons \citep{LeRoCo2015}.} $\Ha \ll 1$, which implies that $\eps=K_2/K_1 \ll 1$, i.e. $m_2 \ll m_1$. 
Making use of this assumption, we obtain simplified expressions for all fitted quantities:
\begin{equation}
\begin{aligned}
        K &=  K_1 (1+\eps \cos \zeta) + \gO(\eps^2,e_k^2,\eps e_k)\, , \\
    \Ha & = - \eps \sin \zeta + \gO(\eps^2,e_k^2,\eps e_k)\, , \\
        c &= k_1+ \gO(\eps^2,e_k^2,\eps e_k)\, , \\
          d & = h_1+ \gO(\eps^2,e_k^2,\eps e_k)\, .
\end{aligned}  
 \label{eq:coef_vr2}
   \end{equation}
All the fitted parameters are directly related to the physical parameters that constrain the orbit of the observed planet, and they additionally provide a simple test for the presence of a co-orbital companion ($\Ha \ne 0$).
For Trojan orbits, $\Ha<0$ (resp. $\Ha>0$) corresponds to the $L4$ (resp. $L5$) point.

\subsection{Anti-transit information}
\label{sec:ATI}

Whenever it is possible to observe the secondary eclipse of the transiting planet at a time $t=t_a$, we can access directly the quantity $k_1$ by comparing the duration between the primary and secondary transit to half the orbital period computed from the two primary transits \citep{Bi1960}
\begin{equation}
k_1 = \frac14 (n t_a - \pi) + \gO(e^2) \, . \label{eq:antitransit}
\end{equation}
In this case, since we can get the $c=k_1$ parameter from the secondary eclipse (usually with  much greater precision than the radial-velocity measurements), we can fix it in expression (\ref{eq:RVf}), and thus fit the only four remaining parameters.
This allows us to achieve a better precision for $\Ha$, and thus confirm the presence of a co-orbital companion.
%

\subsection{Duration of the transits}

The observation of the secondary eclipse of the transiting planet can also constrain the quantity $h_1$ by comparing the duration of the primary transit and the secondary eclipse, $\Delta t$ and $\Delta t_a$, respectively. We have \citep{Bi1960}:
\begin{equation}
h_1 = \frac{\Delta t-\Delta t_a}{\Delta t+\Delta t_a} + \gO(e^2) \, . \label{eq:transitduration}
\end{equation}
In this case, we also get an estimation for the  $d=h_1$ parameter before the fit, which can further improve the determination of $\Ha$. We note, however, that unlike for $k_1$, the precision of this term is not necessarily better than the radial-velocity constrain \citep{MaWi2009}.

\section{False positives}
\label{sec:FP}

There are other physical effects that can also provide non-zero $\Ha$, and thus eventually mimic the presence of a co-orbital companion.
The main sources of error could be due to non-spherical gravitational potentials, the presence of orbital companions, or the presence of an exomoon.

The main consequence of most of the perturbations (general relativity, the $J_2$ of the star and/or of the planet, tidal deformation of the star and/or of the planet, secular gravitational interactions with other planetary companions) is in the precession rate of the argument of the pericentre, $\dot \omega$. 
However, the mean motion frequency that is determined using the radial-velocity and the transits technique is given by $n = \dot \lambda$ (Eq.\,\ref{eq:longitude}), which already contains $\dot \omega$.
Thus, in all these cases our method is still valid.

For close-in companions, $\dot \omega$ cannot be considered  constant, and we can observe a non-zero $\Ha$ value that could mimic the presence of a co-orbital companion.
However, strong interactions require large mass companions whose trace would be independently detected in the radial-velocity data and through TTVs.
The only exceptions are exomoons, which have the exact same mean motion frequency as the observed planet, or the 2:1 mean-motion resonances with small eccentricity, whose harmonics of the radial-velocity data coincide with the co-orbital values.

In the case of exomoons, the satellite switches its orbital position with the planet rapidly, so $\Ha$ oscillates around zero with a frequency $\nu \sim n$ that is not compatible with a libration frequency of a co-orbital companion.
For most of co-orbital configurations, the libration frequency is comparable with the libration frequency at the $L_4$ equilibrium, $\nu=n \sqrt{27/4(m_1+m_2)/m_0} \ll n$, and the average of $\Ha$ is around $\zeta = \pm \pi/3$, not zero.
Therefore, our method also provides a tool for detecting exomoons.

For the 2:1 mean-motion resonance, we must distinguish which planet transits. 
If the transiting planet is the inner one, $\Ha$ is impacted by the eccentricity of the outer planet. However, if the outer planet is massive enough to impact the value of $\Ha$, its harmonic of frequency $n/2$ must be visible in the radial-velocity measurement. 
If the transiting planet is the outer one, the inner planet impacts $\Ha$ indirectly by modifying the value of the parameter $c$. 
This is not a problem if this parameter is well constrained by the anti-transit of the transiting planet. 
Moreover, the inner planet would induce TTV on the transiting planet of the order of $m_2/m_0$ \citep{NeVo2014}. 
If the semi-major axis of the transiting planet is not too large, the TTVs should be observed, and here again their frequency allows to distinguish the co-orbital case from the 2:1 resonance.

\section{Conclusion}

In this Letter, we have generalised the method proposed by \citet{FoGa2006} for detecting co-orbital planets with null to moderate eccentricity and any libration amplitude (from the Lagrangian equilibrium to Horseshoe configurations).
For highly eccentric orbits this method is not needed because it is possible to use radial-velocity alone to infer the presence of the co-orbital companion (Eq.\,(\ref{eq:A3})).

Our method is based in only five free parameters that need to be adjusted to the radial-velocity data.
Moreover, when it is also possible to observe the secondary eclipse, we have additional constraints which reduce the number of parameters to adjust.
One of the free parameters, $\Ha$, is simply a measurement for the presence of a co-orbital companion, which is proportional to the mass ratio $m_2/m_1$.
As discussed in section \ref{sec:FP}, other dynamical causes can produce a non-zero $\Ha$. 
However, alternative scenarios would also  significantly impact the TTV and/or the radial-velocity, and allow us to discriminate between them.

Therefore, if $\Ha$ is statistically different from zero and the TTV and radial-velocity do not show any signature of other causes, the observed system is a strong candidate to harbour a co-orbital companion. 
We additionally get an estimation of its mass.
Inversely, if $\Ha$ is compatible with zero, our method rules out a co-orbital companion down to a given mass, provided that $\sin \zeta$ is not too close to zero (Eq.\,(\ref{eq:coef_vr2})).
This is unlikely because $\zeta = 0$ corresponds to a collision between the two planets, and $\zeta = \pi$ can only occur in the Horseshoe configuration, hence when $(m_1+m_2)/m_0 \lesssim 2 \times 10^{-4}$ \citep{LauCha2002}.

\begin{acknowledgements}
The authors acknowledge financial support from the Observatoire de Paris Scientific Council, CIDMA strategic project UID/MAT/04106/2013, and the Marie Curie Actions of the European Commission (FP7-COFUND). Parts of this work have been carried out within the frame of the National Centre for Competence in Research PlanetS supported by the SNSF. 
\end{acknowledgements}

\bibliographystyle{aa}

\end{document}